\documentclass[12pt]{article}
\usepackage{amsmath}
\usepackage{booktabs}
\usepackage{color}
\usepackage{amsfonts}
\usepackage{graphicx}
\usepackage{subfigure}
\usepackage{titlesec}
\usepackage{indentfirst}
\usepackage{graphicx,psfrag,epsf}
\usepackage{enumerate}
\usepackage{natbib}
\usepackage{url} 
\usepackage{algpseudocode}
\usepackage{algorithm}
\usepackage{amsfonts}
\usepackage{multirow}
\usepackage{comment}
\usepackage{siunitx}
\newtheorem{theorem}{Theorem}%
\newtheorem{remark}{Remark}%
\newcommand{\blind}{1}
\newcommand{\norm}[1]{\left\| #1 \right\|}
\newcommand{\cR}{\mathcal{R}}
\begin{document}

\def\spacingset#1{\renewcommand{\baselinestretch}%
{#1}\small\normalsize} \spacingset{1}


\if1\blind
{
  \title{\bf Cyclic-Shift Sparse Kronecker Tensor Classifier for Signal-Region Detection in Neuroimaging}
  \author{Hsin-Hsiung Huang, Yuh-Haur Chen\\
    Department of Statistics and Data Science, University of Central Florida, \\
    Teng Zhang\\
    Department of Mathematics, University of Central Florida}
  \date{}
  \maketitle
} \fi

\if0\blind
{
  \bigskip
  \bigskip
  \bigskip
  \begin{center}
    {\LARGE\bf Alzheimer’s Disease Risk Modeling Using Neuroimaging Data Analysis}
\end{center}
  \medskip
} \fi

\bigskip
\begin{abstract}
This study proposes a cyclic-shift logistic sparse Kronecker product decomposition (SKPD) model for high-dimensional tensor data, enhancing the SKPD framework with a cyclic-shift mechanism for binary classification. The method enables interpretable and scalable analysis of brain MRI data, detecting disease-relevant regions through a structured low-rank factorization. By incorporating a second spatially shifted view of the data, the cyclic-shift logistic SKPD improves robustness to misalignment across subjects, a common challenge in neuroimaging. We provide asymptotic consistency guarantees under a restricted isometry condition adapted to logistic loss. Simulations confirm the model's ability to recover spatial signals under noise and identify optimal patch sizes for factor decomposition. Application to OASIS-1 and ADNI-1 datasets demonstrates that the model achieves strong classification accuracy and localizes estimated coefficients in clinically relevant brain regions, such as the hippocampus. A data-driven slice selection strategy further improves interpretability in 2D projections. The proposed framework offers a principled, interpretable, and computationally efficient tool for neuroimaging-based disease diagnosis, with potential extensions to multi-class settings and more complex transformations.
\end{abstract}
\noindent%

\newpage
\spacingset{1.45} 
\section{Introduction}

\label{sec:intro}
Diagnosing brain diseases such as Alzheimer’s disease (AD) using high-dimensional MRI data remains a critical challenge in modern medicine due to the complexity of spatial patterns and the limitations of imaging technology. Magnetic resonance imaging, while noninvasive, is sensitive to minor movements, reducing usable data and complicating the analysis \citep{Li2016ReviewMRIQuality}. Traditional matrix and tensor regression methods often struggle with computational inefficiency and poor interpretability when applied to large, high-resolution MRI datasets, hampered by their ability to uncover subtle disease signals \citep{zhou2013tensor, Zhou2014RegularizedMatrixRegression}. Moreover, flattening an image into a vector obliterates its spatial correlations, whereas fully Bayesian tensor models—despite their flexibility—often become computationally prohibitive in large‐cohort settings \citep{Li2015SpatialBayesian,lyu2024bayesian}.

The Sparse Kronecker Product Decomposition (SKPD) framework addresses these gaps by preserving tensor structure through Kronecker products, decomposing data into sparse ``location indicators" ($\mathbf{A}$) and the ``dictionaries" ($\mathbf{B}$) components \citep{cai2022kopa,wu2023_sparse}. Building on foundational regularization techniques \citep{Tibshirani1996Lasso}, SKPD offers a scalable, frequentist solution with explicit interpretability, crucial for identifying AD-related brain changes. Compared to convolutional neural networks (CNNs), which excel in prediction but lack transparency, SKPD pinpoints sparse signal regions, aligning with clinical needs for explainable diagnostics \citep{wu2023_sparse}. Advances in logistic matrix regression further support the use of low-rank models for classification tasks, using sparsity and total variation regularization to improve performance \citep{Wang2017GeneralizedScalarImage, zhang2018RankOptimizedLogistic, Hung2019LowRankEstimation}.

We extend SKPD in two directions. First, we embed it in a logistic-regression framework, enabling direct classification of “AD” versus “non-AD’’ while retaining interpretability \citep{Recht2010GuaranteedMinimumRank,Negahban2011LowRankEstimation}. Second, we introduce a \emph{cyclic-shift} augmentation that realigns each scan by small circular shifts before decomposition, markedly improving robustness to the minor mis-registrations endemic to multi-subject MRI; unlike classic periodic-shift models, our approach requires no strict periodicity assumption \citep{Candes2005Decoding}. Together these modifications—building on \citep{wu2023_sparse,huang2024framework}—reduce effective dimensionality and raise classification accuracy, while preserving voxel-level interpretability. The same machinery naturally generalizes to multiclass settings, opening a path toward finer‐grained staging of neurodegenerative disease.

The paper is organized as follows. Section \ref{sec:setup} details the methodology; Section \ref{sec:theory} develops non-asymptotic theory; Section \ref{sec: Simulation experiments} reports simulation studies; Section \ref{sec:mri_application} analyses the OASIS-1 and ADNI-1 datasets; and Section \ref{sec:discussion} concludes with implications and future directions.

\section{Methodology}
\label{sec:setup}
\subsection{Model Setup and Penalized Likelihood}

We consider observations i.i.d. $\{(y_i,\mathbf{X}_i,\mathbf{z}_i)\}_{i=1}^n$ for a binary classification task, where $y_i\in\{0,1\}$ is a label, $\mathbf{X}_i$ is a matrix or tensor of covariates (for example, an image) and $\mathbf{z}_i\in\mathbb{R}^p$ represents additional vector-valued predictors. In the matrix setting, we have $\mathbf{X}_i\in \mathbb{R}^{D_1\times D_2}$ and seek to estimate a coefficient matrix $\mathbf{C}\in \mathbb{R}^{D_1\times D_2}$. The Sparse Kronecker Product Decomposition (SKPD) posits a multi-term factorization:
\begin{equation}
\label{eq: kronecker product}
    \mathbf{C}=
\sum_{r=1}^R \mathbf{A}_r \,\otimes\, \mathbf{B}_r,
\end{equation}
where $\otimes$ is the Kronecker product, each $\mathbf{A}_r\in \mathbb{R}^{p_1\times p_2}$ identifies salient spatial or structural regions (often enforced to be sparse), and $\mathbf{B}_r\in \mathbb{R}^{d_1\times d_2}$ encodes the corresponding shapes or intensities. The dimensions of $\mathbf{A_r}$ and $\mathbf{B_r}$ are satisfy $D_1 = p_1\times d_1$ and $D_2 = p_2\times d_2$. To ensure only a few regions carry signal, we impose the sparsity assumption of the $\mathbf{A_r}$ 
\begin{equation}
\label{ep: sparse assumption}
    \|\mathbf{A}\|_0\le s,\quad 1 \le s \le (p_1p_2).
\end{equation}
This sparsity yields both interpretability and statistical efficiency 
in high-dimensional image analysis.

This can naturally extends to higher-order tensors (e.g., $\mathbf{X}_i\in \mathbb{R}^{D_1\times D_2\times D_3}$, where the dimensions of $\mathbf{A}$ and $\mathbf{B}$ are satisfy $D_1 = p_1\times d_1$, $D_2 = p_2\times d_2$, and $D_3 = p_3\times d_3$). To use the Kronecker product decomposition easier, we define the transformation operator $\mathcal{K}: \mathbb{R}^{(p_1 d_1) \times (p_2 d_2) \times (p_3 d_3)} \to \mathbb{R}^{(p_1 p_2 p_3) \times (d_1 d_2 d_3)}$, where  $\mathbf{C}^{d_1,d_2,d_3}_{k,l,m}$ is the ($k,l,m$)-th block of $\mathbf{C}$ of dimension $d_1\times d_2\times d_3$. This allows us to analyze tensor SKPD using matrix properties by mapping from any tensor $\mathbf{C}$ as:
\begin{equation}
\label{eq:transformation}
    \tilde{\mathbf{C}}_j = \mathcal{K}(\mathbf{C}_j) = \begin{bmatrix} \operatorname{vec}(\mathbf{C}_{j;1,1,1}^{d_1,d_2,d_3}) & \cdots & \operatorname{vec}(\mathbf{C}_{j;p_1,p_2,p_3}^{d_1,d_2,d_3}) \end{bmatrix}^\top = \mathcal{K}(\sum^R_{r=1}\mathbf{A}_r \otimes \mathbf{B}_r)= \sum^R_{r=1}\text{vec}(\mathbf{A}_r)[\text{vec}(\mathbf{B}_r)]^\top,
\end{equation}
where $\mathbf{C}_{i;k,l,m}^{d_1,d_2,d_3}$ denotes the $(k,l,m)$-th block of $\mathbf{X}_i$. This transformation enables matrix-based analysis of tensor SKPD, reducing computational complexity and enhancing interpretability by separating location $\mathbf{A}_r$ and signal characteristics $\mathbf{B}_r$. 

For each observation $i$, the logit of the binary response satisfies
\[
\log\!\biggl(\frac{\Pr(y_i=1 \mid \mathbf{X}_i,\mathbf{z}_i)}{1-\Pr(y_i=1 \mid \mathbf{X}_i,\mathbf{z}_i)}\biggr)
\;=\;
\langle \mathbf{X}_i,\mathbf{C} \rangle
\;+\;
\langle \mathbf{z}_i,\gamma\rangle
\]
where $\mathbf{C}\in \mathbb{R}^{(p_1d_1)\times(p_2d_2)\times(p_3d_3)}$ is a coefficient tensor for the $\mathbf{X}_i$, and $\gamma\in \mathbb{R}^q$ is a coefficient vector that models the effects of the auxiliary covariate data $\mathbf z_i$, such as demographic or clinical variables in MRI studies that influence classification while ensuring interpretability, and $\langle \cdot,\cdot\rangle$ denotes the Frobenius inner product. 

We estimate the parameters $\{\mathbf{A}_r,\mathbf{B}_r\}_{r=1}^R$ and $\gamma$ by minimizing a regularized logistic likelihood. Specifically, let
\begin{equation}
\label{eq: general likelihood function}
    p_i \;=\;\sigma\!\Bigl(\,\langle \mathbf{X}_i, \sum_{r=1}^R \mathbf{A}_r\otimes \mathbf{B}_r \rangle \;+\; \langle \mathbf{z}_i, \gamma\rangle\Bigr)
\end{equation}
where $\sigma(z) = 1/(1+e^{-z})$ is the sigmoid function.

The negative log-likelihood over the $n$ samples is then penalized to encourage sparsity and reduce overfitting in high dimensions:
\begin{equation}
\label{eq:skpd-penalized-loss}
\min_{\{\mathbf{A}_r,\mathbf{B}_r\},\gamma}\;
-\frac{1}{n}\,
\sum_{i=1}^n
\Bigl[
y_i \log(p_i)
\;+\;
(1 - y_i)\log\bigl(1 - p_i\bigr)
\Bigr]
\;+\;
\mathcal{R}\Bigl(\{\mathbf{A}_r,\mathbf{B}_r\},\gamma\Bigr)
\end{equation}
In practice, $\mathcal{R}$ may combine sparsity and smoothness penalties, particularly on $\mathbf{A}_r$ and $\mathbf{B}_r$, to ensure both interpretability and numerical stability.

To explain our choice of $\mathcal{R}$ in \eqref{eq:skpd-penalized-loss}, consider a setting that imposes an $\ell_1$ penalty on each $\mathbf{A}_r$ effectively zeroes out non-essential regions, thus focusing on the most truly discriminative and impactful features, along with $\mathbf{B}_r$ may require a milder regularization, since it captures continuous shape or intensity variations within those identified regions. The elastic net in $\mathbf{B}_r$ gives a mix of variable selection (through $\ell_1$) and stability (through $\ell_2$) when features are correlated. The $\ell_1$ penalty on $\gamma$ ensures sparse covariate effects, critical in high-dimensional settings like medical imaging.
\begin{equation}
   \mathcal{R}\Bigl(\{\mathbf{A}_r,\mathbf{B}_r\},\gamma\Bigr) =  \;\lambda_a \sum_{r=1}^R \|\mathbf{A}_r\|_1 
     \;+\;\lambda_b\!\sum_{r=1}^R \Bigl[\alpha\,\|\mathbf{B}_r\|_1 + (1 - \alpha)\,\|\mathbf{B}_r\|_2\Bigr]+ \lambda_\gamma \|\boldsymbol{\gamma}\|_1
     \label{eq:penalty_term}
\end{equation}
where $\lambda_a, \ \lambda_b, \ \lambda_\gamma>0$ are penalty coefficients, and $\alpha\in[0,1]$ governs the relative weights of $\ell_1$(lasso) and $\ell_2$ (ridge) components in the elastic net.

Overall, the penalized likelihood formula \eqref{eq:penalty_term} not only ensures the model captures essential structures in $\mathbf{C}=\sum_{r=1}^R \mathbf{A}_r\otimes\mathbf{B}_r$ in formula \eqref{eq: kronecker product}, but also controls model complexity so that sparse and interpretable solutions can be achieved in large-scale classification tasks. 

\subsection{Cyclic-Shift Mechanism in SKPD}
\label{sec: spatial shift}

The cyclic-shift mechanism enhances the SKPD framework by improving its ability to handle spatial misalignment and reducing reliance on fixed feature positions. This is achieved by translating the input tensor along its spatial dimensions, generating a second view that complements the original data. This approach is particularly useful in high-dimensional datasets, such as medical imaging (e.g., MRI), where slight positional variations across subjects—due to differences in alignment or preprocessing—can impact feature detection. By applying cyclic-shifts, the model adapts to these variations, ensuring that critical patterns remain detectable regardless of their exact locations, while potentially consolidating signals into fewer regions to reduce the effective number of parameters.

The cyclic-shift operation, inspired by the cyclic-shift matrix (CSM) framework \citep{geng2019cyclic}, translates the input tensor $\mathbf{X}_i \in \mathbb{R}^{(p_1 d_1) \times (p_2 d_2) \times (p_3 d_3)}$ by a fixed offset along each spatial dimension. Initially, the unshifted tensor is transformed using the Kronecker structure mapping $\mathcal{K}$ from \eqref{eq:transformation}, resulting in $\tilde{\mathbf{X}}_{1,i} = \mathcal{K}(\mathbf{X}_i) \in \mathbb{R}^{p \times d}$, where $p = p_1 p_2 p_3$ and $d = d_1 d_2 d_3$. A shifted version, $\tilde{\mathbf{X}}_{2,i}$, is then created by translating the tensor indices before transformation. Specifically, we construct the shifted tensor $\mathbf{X}_i^{\text{shifted}}$ using cyclic-shifts based on the grid cells defined by the SKPD coefficient matrix $\mathbf{A}_{v,r} \in \mathbb{R}^{p_1 \times p_2 \times p_3}$. Each cell in the grid corresponds to a patch in $\mathbf{X}_i$ of size $(d_1, d_2, d_3)$, computed as the ratio of the tensor dimensions to the grid dimensions, i.e., $d_1 = (p_1 d_1)/p_1$, $d_2 = (p_2 d_2)/p_2$, $d_3 = (p_3 d_3)/p_3$. The shift offsets are set to half the cell size to ensure a minor adjustment, i.e., $s_1 = d_1/2$, $s_2 = d_2/2$, $s_3 = d_3/2$, applied along each dimension.

The cyclic-shift is implemented using CSMs $\mathbf{Q}_1 \in \mathbb{R}^{(p_1 d_1) \times (p_1 d_1)}$ and $\mathbf{Q}_2 \in \mathbb{R}^{(p_2 d_2) \times (p_2 d_2)}$, where each $\mathbf{Q}_k$ has ones on its superdiagonal and the bottom-left corner, with zeros elsewhere, enabling a cyclic-shift per application. The shifted tensor is defined as:
\begin{equation}
\label{eq:cyclic_shift_tensor}
\mathbf{X}_i^{\text{shifted}} = (\mathbf{Q}^{\top})^{s_1} \mathbf{X}_i \mathbf{Q}^{s_2},
\end{equation}
where $(\mathbf{Q}^{\mathrm{T}})^{s_1}$ shifts the rows downward by $s_1$ pixels and $\mathbf{Q}^{s_2}$ shifts the columns rightward by $s_2$ pixels, all with cyclic wrap-around for out-of-bounds indices. Since $\mathcal{K}$ in \eqref{eq:transformation} reshapes the tensor into a matrix, this operation effectively translates the indices of the resulting matrix, though it is applied in the tensor space. This formulation can be extended to higher dimensions by including additional CSMs for other axes, e.g., a third dimension with a corresponding shift. The shifted tensor is then transformed to $\tilde{\mathbf{X}}_{2,i} = \mathcal{K}(\mathbf{X}_i^{\text{shifted}}) \in \mathbb{R}^{p \times d}$, which is paired with the corresponding coefficient tensor $\mathbf{C}_2$.

\begin{figure}[ht]
    \centering
    \includegraphics[width=0.8\linewidth]{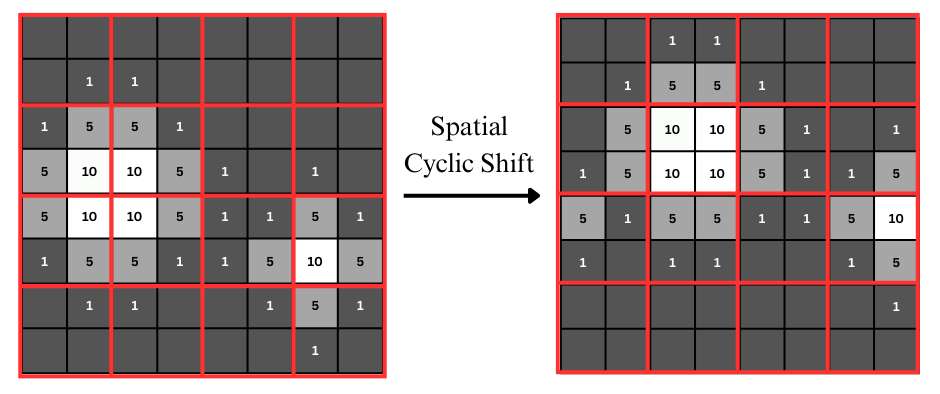}
    \caption{Illustration of the spatial cyclic-shift mechanism on a $4 \times 4$ grid (outlined in red) for the coefficient matrix $\mathbf{A}$. In the initial arrangement (left), a large signal region extends across several neighboring cells, while a smaller signal block occupies a single cell. After a half-cell cyclic shift in both horizontal and vertical directions (right), the large signal is concentrated into one cell, reducing the number of active regions required for representation.}
    \label{fig:spatial_shift_on_X}
\end{figure}

Figure \ref{fig:spatial_shift_on_X} illustrates this process on a $4\times4$ grid, where each cell represents a $2\times2$ pixel region (for an $8\times8$ image). In the original view (left), a large signal region spans multiple cells, and a smaller signal block occupies a single cell, resulting in multiple active regions. After a cyclic-shift of the cell size—in both dimensions (right), the large signal consolidates into a single cell, reducing the total active regions. This example demonstrates how spatial shifting can concentrate signals, potentially reducing the number of parameters needed for representation by requiring fewer non-zero entries in the coefficient matrices.

To integrate this mechanism into the SKPD model, two sets of coefficient matrices are introduced: $\mathbf{C}_1 = \sum_{r=1}^R \mathbf{A}_{1,r} \otimes \mathbf{B}_{1,r}$ and $\mathbf{C}_2 = \sum_{r=1}^R \mathbf{A}_{2,r} \otimes \mathbf{B}_{2,r}$, each paired with its respective transformed tensor $\tilde{\mathbf{X}}_{1,i}$ and $\tilde{\mathbf{X}}_{2,i}$. The factor matrices $\mathbf{A}_{2,r}$ and $\mathbf{B}_{2,r}$ are learned based on the shifted input $\tilde{\mathbf{X}}_{2,i}$, but the inner product $\langle \tilde{\mathbf{X}}_{2,i}, \mathbf{C}_2 \rangle$ is constructed to correspond to the original spatial coordinates, effectively aligning the contribution of the shifted view back to the unshifted space. This alignment ensures that combining the contributions from both views captures complementary spatial information without misalignment. During training, optimization alternates between these coefficient sets, ensuring robust feature extraction without overfitting to specific locations.

The final likelihood combines contributions from both views, along with an offset term $\mathbf{z}_i$ for additional covariates, as:
\begin{equation}
    \label{eq: overall likelihood}
    p_i = \sigma\left( \langle \tilde{\mathbf{X}}_{1,i}, \mathbf{C}_1 \rangle + \langle \tilde{\mathbf{X}}_{2,i}, \mathbf{C}_2 \rangle + \langle \mathbf{z}_i, \boldsymbol{\gamma} \rangle \right)
\end{equation}

This structured approach enhances the model’s ability to disentangle sparse or overlapping signals, improving both interpretability and robustness. Overall, the cyclic-shift mechanism allows the SKPD model to generalize effectively across varying subject alignments, making it well-suited for medical imaging applications where spatial consistency is often imperfect.

\subsection{A path following parameter estimation algorithm}
\label{sec: algorithm}

To simplify the tensor inner product computation in the SKPD model, we transform the Kronecker product into a scalar form using vectorization. For view $ v $ (where $ v = 1, 2 $), the inner product is:
\[
\langle \tilde{\mathbf{X}}_{v,i}, \mathbf{C}_v \rangle = \left\langle \tilde{\mathbf{X}}_{v,i}, \sum_{r=1}^{R} \mathbf{A}_{v,r} \otimes \mathbf{B}_{v,r} \right\rangle = \sum_{r=1}^{R} \text{vec}(\mathbf{A}_{v,r})^\top \tilde{\mathbf{X}}_{v,i} \text{vec}(\mathbf{B}_{v,r}),
\]
where $\tilde{\mathbf{X}}_{v,i} = \mathcal{K}(\mathbf{X}_{v,i}) \in \mathbb{R}^{p \times d}$ is the transformed input tensor for view $ v $, with $ p = p_1 p_2 p_3 $, $ d = d_1 d_2 d_3 $, via the operator $\mathcal{K}$ defined in \eqref{eq:transformation}, $\text{vec}(\mathbf{A}_{v,r}) \in \mathbb{R}^p$, and $\text{vec}(\mathbf{B}_{v,r}) \in \mathbb{R}^d$. This leverages the property $\mathcal{K}(\mathbf{A}_{v,r} \otimes \mathbf{B}_{v,r}) = \text{vec}(\mathbf{A}_{v,r}) [\text{vec}(\mathbf{B}_{v,r})]^\top$, reducing computational complexity and aligning tensor operations with matrix-based optimization. The estimator $(\hat{\mathbf{C}}, \hat{\boldsymbol{\gamma}})$ is obtained by alternately optimizing $\mathbf{A}_{v,r}$ and $\mathbf{B}_{v,r}$ to minimize the optimization problem in \eqref{eq:penalty_term}, with $\boldsymbol{\gamma} \in \mathbb{R}^q$ as the covariate coefficient.

We initialize $\hat{\mathbf{A}}_{v,r}^{(0)}$ as the top-$ R $ left singular vectors of the weighted average transformed data $\sum_{i=1}^n y_i \tilde{\mathbf{X}}_{v,i}$, capturing the principal directions of the signal associated with the response $ y_i $. The parameters $\hat{\mathbf{B}}_{v,r}^{(0)}$ are set to vectors of ones, and $\hat{\boldsymbol{\gamma}}^{(0)}$ is initialized to zeros (or a preliminary estimate if covariates $\mathbf{z}_i$ are present). The updates for $\mathbf{B}_{v,r}$, $\mathbf{A}_{v,r}$, and $\boldsymbol{\gamma}$ minimize the penalized logistic loss, leveraging the transformed inputs and offsets to isolate each parameter's contribution. Below, we detail the mathematical formulations in vectorized notation.

Given fixed $\hat{\mathbf{A}}_{v,r}^{(t)}$, $\hat{\mathbf{A}}_{v',r}^{(t)}$, $\hat{\mathbf{B}}_{v',r}^{(t)}$ (where $ v' \neq v $), and $\hat{\boldsymbol{\gamma}}^{(t)}$, we update $\mathbf{B}_{v,r}$ as:
\begin{align*}
     \hat{\mathbf{B}}_{v,r}^{(t+1)} &= \arg\min_{\mathbf{B}_{v,1}, \dots, \mathbf{B}_{v,R}} \left\{ -\frac{1}{n} \sum_{i=1}^n [y_i \log(\hat{p}_i) + (1 - y_i) \log(1 - \hat{p}_i)] + \lambda_b \sum_{r=1}^R \left( \alpha \|\mathbf{B}_{v,r}\|_1 + (1 - \alpha) \|\mathbf{B}_{v,r}\|_F^2 \right) \right\} \\
     \hat{p}_i &= \sigma \left( \sum_{r=1}^R \text{vec}(\hat{\mathbf{A}}_{v,r}^{(t)})^\top \tilde{\mathbf{X}}_{v,i} \text{vec}(\mathbf{B}_{v,r}) + \text{offset}_{v,i} \right)
\end{align*}
where the offset isolates the fixed terms:
    \[
    \text{offset}_{v,i} = \left\langle \tilde{\mathbf{X}}_{v',i}, \sum_{r=1}^R \hat{\mathbf{A}}_{v',r}^{(t)} \otimes \hat{\mathbf{B}}_{v',r}^{(t)} \right\rangle + \langle \mathbf{z}_i, \hat{\boldsymbol{\gamma}}^{(t)} \rangle
    \]

    Define the feature matrix for optimization:
    \[
    \mathbf{F}_{v} = \begin{bmatrix}
    \left( \sum_{r=1}^R \tilde{\mathbf{X}}_{v,1}^\top \text{vec}(\hat{\mathbf{A}}_{v,r}^{(t)}) \right)^\top \\
    \vdots \\
    \left( \sum_{r=1}^R \tilde{\mathbf{X}}_{v,n}^\top \text{vec}(\hat{\mathbf{A}}_{v,r}^{(t)}) \right)^\top
    \end{bmatrix} \in \mathbb{R}^{n \times (d R)}
    \]
    where each row stacks the contributions of $\tilde{\mathbf{X}}_{v,i} \hat{\mathbf{A}}_{v,r}^{(t)}$ across $ r $. The linear predictor becomes $\mathbf{F}_v \text{vec}(\mathbf{B}_v) + \text{offset}_v$, with $\text{vec}(\mathbf{B}_v) = [\text{vec}(\mathbf{B}_{v,1})^\top, \dots, \text{vec}(\mathbf{B}_{v,R})^\top]^\top$.

    Similarly, for updating $\mathbf{A}_{v,r}$ given the updated $\hat{\mathbf{B}}_{v,r}^{(t+1)}$, fixed $\hat{\mathbf{A}}_{v',r}^{(t)}$, $\hat{\mathbf{B}}_{v',r}^{(t+1)}$, and $\hat{\boldsymbol{\gamma}}^{(t)}$, we solve:
    \begin{align*}
         \hat{\mathbf{A}}_{v,r}^{(t+1)} &= \arg\min_{\mathbf{A}_{v,1}, \dots, \mathbf{A}_{v,R}} \left\{ -\frac{1}{n} \sum_{i=1}^n [y_i \log(\hat{p}_i) + (1 - y_i) \log(1 - \hat{p}_i)] + \lambda_a \sum_{r=1}^R \|\mathbf{A}_{v,r}\|_1 \right\} \\
         \hat{p}_i &= \sigma \left( \sum_{r=1}^R \text{vec}(\mathbf{A}_{v,r})^\top \tilde{\mathbf{X}}_{v,i} \text{vec}(\hat{\mathbf{B}}_{v,r}^{(t+1)}) + \text{offset}_{v,i} \right) \\
         \text{offset}_{v,i} &= \left\langle \tilde{\mathbf{X}}_{v',i}, \sum_{r=1}^R \hat{\mathbf{A}}_{v',r}^{(t)} \otimes \hat{\mathbf{B}}_{v',r}^{(t+1)} \right\rangle + \langle \mathbf{z}_i, \hat{\boldsymbol{\gamma}}^{(t)} \rangle
    \end{align*}

    Define the feature matrix:
    \[
    \mathbf{G}_v = \begin{bmatrix}
    \left( \sum_{r=1}^R \tilde{\mathbf{X}}_{v,1} \text{vec}(\hat{\mathbf{B}}_{v,r}^{(t+1)}) \right)^\top \\
    \vdots \\
    \left( \sum_{r=1}^R \tilde{\mathbf{X}}_{v,n} \text{vec}(\hat{\mathbf{B}}_{v,r}^{(t+1)}) \right)^\top
    \end{bmatrix} \in \mathbb{R}^{n \times (p R)}
    \]
    where each row stacks $\tilde{\mathbf{X}}_{v,i}^\top \text{vec}(\hat{\mathbf{B}}_{v,r}^{(t+1)})$ across $ r $. The linear predictor is $\mathbf{G}_v \text{vec}(\mathbf{A}_v) + \text{offset}_v$, with $\text{vec}(\mathbf{A}_v) = [\text{vec}(\mathbf{A}_{v,1})^\top, \dots, \text{vec}(\mathbf{A}_{v,R})^\top]^\top$.
    
    Finally, update the coefficient $\boldsymbol{\gamma}$ with the updated image coefficients:
    \begin{align*}
         \hat{\boldsymbol{\gamma}}^{(t+1)} &= \arg\min_{\boldsymbol{\gamma}} \left\{ -\frac{1}{n} \sum_{i=1}^n [y_i \log(\hat{p}_i) + (1 - y_i) \log(1 - \hat{p}_i)] + \lambda_\gamma \|\boldsymbol{\gamma}\|_1 \right\} \\
         \hat{p}_i &= \sigma \left( \sum_{v=1}^2 \left\langle \tilde{\mathbf{X}}_{v,i}, \sum_{r=1}^R \hat{\mathbf{A}}_{v,r}^{(t+1)} \otimes \hat{\mathbf{B}}_{v,r}^{(t+1)} \right\rangle + \mathbf{z}_i^\top \boldsymbol{\gamma} \right)
    \end{align*}
    where the linear predictor is $\mathbf{z}_i^\top \boldsymbol{\gamma} + \text{offset}_i$, with $\text{offset}_i = \sum_{v=1}^2 \left\langle \tilde{\mathbf{X}}_{v,i}, \sum_{r=1}^R \hat{\mathbf{A}}_{v,r}^{(t+1)} \otimes \hat{\mathbf{B}}_{v,r}^{(t+1)} \right\rangle$.

    \begin{algorithm}[ht]
\caption{Cyclic-Shift Logistic SKPD with Covariate Adjustment}
\begin{algorithmic}[1]
  \State \textbf{Input:} $ y_i $, $ \mathbf{X}_i $, $ \mathbf{z}_i $, $ i = 1, \dots, n $; $ \lambda_a, \lambda_b, \lambda_\gamma $; $ R $; $ T $
  \State \textbf{Initialize:} For $ v = 1, 2 $, $ \hat{\mathbf{A}}_{v,r}^{(0)} $: top-$ R $ left singular vectors of $ \sum_{i=1}^n y_i \tilde{\mathbf{X}}_{v,i} $, $ \tilde{\mathbf{X}}_{v,i} = \mathcal{K}(\mathbf{X}_{v,i}) $; $ \hat{\mathbf{B}}_{v,r}^{(0)} = \mathbf{1} $; $ \hat{\boldsymbol{\gamma}}^{(0)} = \mathbf{0} $
  \For{$ t = 0, 1, \dots, T-1 $}
    \For{$ v = 1, 2 $}
      \State \textbf{Update $ \hat{\mathbf{B}}_{v,r}^{(t+1)} $:} Minimize loss with $ \mathbf{F}_v = [\tilde{\mathbf{X}}_{v,i}^\top \text{vec}(\hat{\mathbf{A}}_{v,r}^{(t)})]_{i=1}^n $, offset $ \langle \tilde{\mathbf{X}}_{v',i}, \sum_r \hat{\mathbf{A}}_{v',r}^{(t)} \otimes \hat{\mathbf{B}}_{v',r}^{(t)} \rangle + \langle \mathbf{z}_i, \hat{\boldsymbol{\gamma}}^{(t)} \rangle $, $ v' \neq v $
      \State \textbf{Update $ \hat{\mathbf{A}}_{v,r}^{(t+1)} $:} Minimize loss with $ \mathbf{G}_v = [\tilde{\mathbf{X}}_{v,i} \text{vec}(\hat{\mathbf{B}}_{v,r}^{(t+1)})]_{i=1}^n $, offset $ \langle \tilde{\mathbf{X}}_{v',i}, \sum_r \hat{\mathbf{A}}_{v',r}^{(t)} \otimes \hat{\mathbf{B}}_{v',r}^{(t+1)} \rangle + \langle \mathbf{z}_i, \hat{\boldsymbol{\gamma}}^{(t)} \rangle $, $ v' \neq v $
    \EndFor
    \State \textbf{Update $ \hat{\boldsymbol{\gamma}}^{(t+1)} $:} Minimize loss with $ \mathbf{Z} = [\mathbf{z}_1^\top, \dots, \mathbf{z}_n^\top]^\top $, offset $ \sum_{v=1}^2 \langle \tilde{\mathbf{X}}_{v,i}, \sum_r \hat{\mathbf{A}}_{v,r}^{(t+1)} \otimes \hat{\mathbf{B}}_{v,r}^{(t+1)} \rangle $
    \State \textbf{Check Convergence}
  \EndFor
  \State \textbf{Return:} $ \hat{\mathbf{A}}_{v,r}^{(T)} $, $ \hat{\mathbf{B}}_{v,r}^{(T)} $, $ \hat{\boldsymbol{\gamma}}^{(T)} $
\end{algorithmic}
\end{algorithm}

\section{Theoretical Results}
\label{sec:theory}

We provide theoretical justification for the cyclic-shift SKPD framework, extending the original SKPD regression model from \citep{wu2023_sparse} to binary classification using a logistic model. Our analysis focuses on third-order tensors $\mathbf{X}_i \in \mathbb{R}^{D_1 \times D_2 \times D_3}$ that may exhibit spatial misalignments, where $D_k = p_k \times d_k$ for $k=1,2,3$, with $p = p_1 p_2 p_3$ and $d = d_1 d_2 d_3$ denoting the dimensions of the Kronecker factors. This formulation leverages the spatial structures inherent in medical imaging data, with natural extensions to two-dimensional matrices (as a special case) or higher-order tensors for more complex applications.

Our key contributions include adapting the SKPD framework to a logistic loss function for binary classification and introducing a cyclic-shift mechanism to handle misaligned tensor inputs. Section~\ref{sec:assumptions} outlines key assumptions, including a restricted isometry property (RIP) tailored for the logistic model and initialization proximity, which ensure identifiability and well-conditioning for consistent estimation of the true coefficient tensor $\mathbf{C}^*$ and covariate parameters $\boldsymbol{\gamma}^*$. 

Sections~\ref{sec:consistency-skpd} present consistency results under these assumptions, showing that $\mathbf{C}$ can be estimated at a rate on the order of $\sqrt{(\#\text{parameters})/n}$, where the parameter count reflects the low-rank and sparse structure enhanced by the cyclic-shift approach.

\subsection{Key Assumptions}
\label{sec:assumptions}

A central challenge in high-dimensional logistic regression with tensor covariates is to ensure the data retains enough information to estimate the sparse, Kronecker-structured coefficients $\mathbf{C}$. We require conditions analogous to the restricted isometry property (RIP), which ensures that no low-complexity structure is severely distorted by the design. Here we only mentioned the key assumptions for the theorem, More comprehensive assumption description can be find in supplementary material.

Let $\Pi\colon \mathbb{R}^{D_1\times \cdots \times D_k}\to\mathbb{R}^n$ be the mapping induced by the data $\{\mathbf{X}_i\}$, i.e., $\Pi(\mathbf{C})$ collects the scalar products $\{\langle \mathbf{X}_i,\mathbf{C}\rangle\}$. A rank-based or Kronecker-based RIP requires that $\Pi$ behave near-isometrically on tensors with up to $R$ Kronecker products terms, corresponding to a manifold of dimension $\sum_{r=1}^R \dim(\mathbf{A}_r,\mathbf{B}_r)$. When such an RIP holds, for any tensor $\mathbf{C} = \sum_{r=1}^R \mathbf{A}_r \otimes \mathbf{B}_r$ with $\|\mathbf{A}_r\|_0 \leq 2 s_r$, there exists a constant $\delta_{2R} \in (0, 1)$ such that:
\begin{equation}
\label{eq:assumption_RIP}
(1 - \delta_{2R}) \|\mathbf{C}\|_F^2 \leq \frac{1}{n} \sum_{i=1}^n \langle \tilde{\mathbf{X}}_i, \mathcal{K}(\mathbf{C}) \rangle^2 \leq (1 + \delta_{2R}) \|\mathbf{C}\|_F^2,
\end{equation}
where $\tilde{\mathbf{X}}_i = \mathcal{K}(\mathbf{X}_i) \in \mathbb{R}^{p\times d}$ is the transformed design, and $\mathcal{K}$ maps the tensor to a matrix representation, shown in equation~\eqref{eq:transformation}. This RIP ensures the logistic loss function remains well-conditioned locally, preventing degeneracies in gradient-based updates and enabling consistent estimation.

Additionally, we need a good initialization to converge to the true parameters. We assume the initial estimates satisfy:
\begin{equation}
\label{eq:assumption_bounded_design}
\sum_{r=1}^R \|\hat{\mathbf{A}}_r^{(0)} - \mathbf{A}_r^*\|_F^2 \leq \eta_0, \quad \sum_{r=1}^R \|\hat{\mathbf{B}}_r^{(0)} - \mathbf{B}_r^*\|_F^2 \leq \eta_0, \quad \|\hat{\gamma}^{(0)} - \gamma^*\|_2^2 \leq \eta_0,
\end{equation}
for a small constant $\eta_0 > 0$, typically achieved via spectral methods (e.g., singular value decomposition of a gradient-based surrogate).

Lastly, the sample size must be sufficient to estimate the model’s parameters, scaling with the effective degrees of freedom. We require:
\begin{equation}
\label{eq:assumption_sample}
n \geq c \left( R (p+d) \log(p) + q \right),
\end{equation}
where $c > 0$ is a constant, $p$ and $d$ are the dimensions of the transformed tensor space, and $q$ is the dimension of the covariate vector $\mathbf{z}_i$. More assumptions are listed in the supplementary material from equation~\eqref{eq:assumption_bounded_design} to equation~\eqref{eq:assumption_sample}.   These conditions collectively ensure the SKPD estimator’s consistency, as established in the following theorem.

\subsection{Consistency of the Cyclic-Shift Logistic SKPD Estimator}
\label{sec:consistency-skpd}

Consider a logistic regression model with a cyclic-shift SKPD estimator for binary classification, utilizing two views: the original tensor input and its spatially shifted counterpart. The estimator applies an $\ell_1$-penalty on $\mathbf{A}_{v,r}$ and an Elastic Net penalty on $\mathbf{B}_{v,r}$. For each sample $i$, the inputs are $\tilde{\mathbf{X}}_{1,i} = \mathcal{K}(\mathbf{X}_{1,i})$ (original) and $\tilde{\mathbf{X}}_{2,i} = \mathcal{K}(\mathbf{X}_{\text{shifted},i})$ (shifted), where $\mathcal{K}$ is the Kronecker structure mapping defined in equation~\eqref{eq:transformation}. The indices $v=1,2$ denote the original and shifted views, respectively. The estimator and true coefficient tensors are:
\begin{equation*}
\hat{\mathbf{C}} = \sum_{v=1}^2 \hat{\mathbf{C}}_v = \sum_{v=1}^2 \sum_{r=1}^R \hat{\mathbf{A}}_{v,r} \otimes \hat{\mathbf{B}}_{v,r},
\end{equation*}
\begin{equation*}
\mathbf{C}^* = \sum_{v=1}^2 \mathbf{C}_v^* = \sum_{v=1}^2 \sum_{r=1}^R \mathbf{A}_{v,r}^* \otimes \mathbf{B}_{v,r}^*,
\end{equation*}
where $\hat{\mathbf{A}}_{v,r}, \mathbf{A}_{v,r}^* \in \mathbb{R}^{p_1 \times p_2 \times p_3}$, $\hat{\mathbf{B}}_{v,r}, \mathbf{B}_{v,r}^* \in \mathbb{R}^{d_1 \times d_2 \times d_3}$, $p = p_1 p_2 p_3$, $d = d_1 d_2 d_3$, and $\otimes$ is the tensor Kronecker product. Let $\boldsymbol{\gamma} \in \mathbb{R}^{q}$ collect additional linear covariates and write $f \asymp g$ if $C' g \leq f \leq C g$ for some constants $C, C' > 0$.

\begin{theorem}[Consistency of the Cyclic-Shift Logistic SKPD Estimator]
\label{thm:consistency}
Under the assumptions listed in the supplementary material, suppose the design satisfies a Joint Restricted Isometry Property (RIP) to ensure the combined tensor structure is well-conditioned. For a universal constant $c > 0$, assume:
\begin{equation}
\label{eq:assumption_consistency}
n \geq c \left( 2 R (p + d) \log p + q \right).
\end{equation}
Then, with probability at least $1 - c_1 \exp(-c_2 n) - c_3 \exp(-c_4 (2 R (p + d) + q))$, the estimators $\hat{\mathbf{C}}$ and $\hat{\boldsymbol{\gamma}}$ satisfy:
\begin{equation}
\label{eq:bound_consistency}
\norm{\hat{\mathbf{C}} - \mathbf{C}^*}_F + \norm{\hat{\boldsymbol{\gamma}} - \boldsymbol{\gamma}^*}_2 \asymp \sqrt{\frac{R (p + d) \log p + q}{n}} + \sqrt{\frac{\cR(\mathbf{C}^*, \boldsymbol{\gamma}^*)}{n}},
\end{equation}
where $C = \max(c_0, c_3, c_5, c_6)$, $C' = \min(c_1, c_2, c_3, c_4)$, the sparsity-inducing penalty is defined in~\eqref{eq:penalty_term}, and $\lambda_a, \lambda_b, \lambda_\gamma \propto \sqrt{\frac{\log p}{n}}$.
\end{theorem}

\begin{remark}
The cyclic-shift mechanism enhances the logistic SKPD estimator by leveraging the structural alignment between the original and shifted views, potentially reducing the effective number of parameters needed to represent $\mathbf{C}^*$. Suppose the total sparsity or magnitude of coefficients across both views satisfies:
\[
\cR_{\text{shift}}(\mathbf{C}^*, \boldsymbol{\gamma}^*) \leq \cR(\mathbf{C}, \boldsymbol{\gamma}^*) - \Delta,
\]
for some $\Delta > 0$, where $\cR_{\text{shift}}$ is the penalty for the cyclic-shift model with $2R$ terms, and $\cR$ is the penalty for a single-view SKPD model with $R$ terms, defined as $\mathbf{C} = \sum_{r=1}^R \mathbf{A}_r \otimes \mathbf{B}_r$. Then, the cyclic-shift estimator achieves a tighter error bound:
\[
\norm{\hat{\mathbf{C}}_{\text{shift}} - \mathbf{C}^*}_F + \norm{\hat{\boldsymbol{\gamma}} - \boldsymbol{\gamma}^*}_2 < \norm{\hat{\mathbf{C}} - \mathbf{C}^*}_F + \norm{\hat{\boldsymbol{\gamma}} - \boldsymbol{\gamma}^*}_2.
\]
This reduction in effective parameters enhances estimation efficiency.
\end{remark}

This theorem establishes the consistency of the cyclic-shift logistic SKPD estimator, extending the SKPD framework from linear regression \citep{wu2023_sparse} to binary classification. By integrating two views---original and spatially shifted tensor inputs---within a logistic regression framework, and using alternating optimization of $\mathbf{A}_{v,r}$ and $\mathbf{B}_{v,r}$, the approach leverages a Joint Restricted Isometry Property to ensure a well-conditioned design. The cyclic-shift mechanism improves efficiency by aligning features across views, leading to a tighter error bound. The proof, provided in the supplementary material, shows that, under a mild sample-size requirement, the cyclic-shift logistic SKPD estimator attains the near-optimal rate $\asymp \sqrt{\frac{R (p + d) \log p + q}{n}} + \sqrt{\frac{\cR(\mathbf{C}^*, \boldsymbol{\gamma}^*)}{n}}$. The theorem extends the SKPD consistency result of \citep{wu2023_sparse} from linear to logistic regression.

\subsection{Properties of the Cyclic-Shifted SKPD Estimator}
We present an example where the true coefficient tensor $\mathbf{C}^*$ is generated by the cyclic-shifted SKPD model using the second view (spatially shifted input) with rank $R=1$. However, the standard SKPD model cannot accurately recover $\mathbf{C}^*$ unless $R \geq 4$.

\begin{theorem}
Suppose the true parameter $\mathbf{C}^*$ is generated by the cyclic-shifted model
\[
\mathbf{C}^* = \sum_{r=1}^R \hat{\mathbf{A}}_{2,r} \otimes \hat{\mathbf{B}}_{2,r},
\]
where $R = 1$, $\hat{\mathbf{A}}_{2,1}$ is a matrix with exactly one nonzero element, and $\hat{\mathbf{B}}_{2,1}$ is a matrix with all entries equal to $1$. Then, the standard SKPD estimator with $R < 4$ is inconsistent, in the sense that
\[
\min_{\{\mathbf{A}_r, \mathbf{B}_r\}_{r=1}^R} \left\| \sum_{r=1}^R \mathbf{A}_r \otimes \mathbf{B}_r - \mathbf{C}^* \right\|_F \geq \sqrt{\frac{4-R}{4}} \left\| \mathbf{C}^* \right\|_F.
\]
\end{theorem}

\begin{remark}[Interpretation]
When the regularization parameter $\lambda = 0$ or is bounded by a constant, the dominant error term is $\sqrt{\frac{R (p + d) \log(p) + q}{n}}$, analogous to rank-$R$ terms in low-rank matrix regression but adapted to the tensor structure and logistic loss. The $\log(p)$ term accounts for the nonlinearity of the logistic model, while the cyclic-shift mechanism improves efficiency by reducing the effective number of parameters. In practice, $R$ is typically small, and $\mathbf{A}_r$ and $\mathbf{B}_r$ are sparse, resulting in a model with significantly fewer parameters than an unstructured $\mathbf{C}$. This makes the model well-suited for high-dimensional tensor data. Ongoing work focuses on deriving tighter error bounds that incorporate rank constraints, sparsity, and potential block structures in $\mathbf{B}_{v,r}$, as well as exploring minimax optimality under specific assumptions on $\mathbf{X}_i$. The Kronecker-based SKPD framework thus provides a flexible and computationally efficient approach for high-dimensional classification with tensor covariates.
\end{remark}

\section{Simulation Study}
\label{sec: Simulation experiments}
In this section, we conduct simulation studies to evaluate the proposed cyclic-shift SKPD logistic model’s performance in region detection and classification under controlled conditions. We perform three experiments: (1) a 2D matrix simulation to assess hyperparameter tuning and patch size effects on signal reconstruction, motivated by the need to optimize model settings for MRI-like data; (2) a noise robustness study to test the model’s limits and the cyclic-shift’s effectiveness, critical for handling noisy neuroimaging data; and (3) a 3D tensor simulation to validate the model’s ability to capture complex spatial patterns, mirroring real-world MRI structures. These experiments use synthetic datasets with predefined shapes (e.g., butterfly, circles for 2D, and two balls for 3D) and varying noise levels, providing insights into the model’s scalability, robustness, and applicability to MRI signal detection.

\subsection{Simulation Settings}

To test the SKPD model’s performance, we create synthetic datasets using 128x128 grayscale shape templates (e.g., butterfly, circles), normalized to $[0,1]$, where $\mathbf{C}_{\text{signal}} \in \mathbb{R}^{128\times 128}$ is the true signal pattern (extended to 3D tensors like two-ball shapes in later experiments). Each dataset splits evenly into two classes: signal-plus-noise ($Y=1$), where $X_i = \mathbf{C}_{\text{signal}} + \text{noise}$, and noise-only ($Y=0$), where $X_i = \text{noise}$, with noise drawn from a Gaussian distribution. A response, $\langle X_i, \mathbf{C}_{\text{signal}} \rangle + \epsilon$ (Gaussian noise), is computed and transformed via a sigmoid function to assign probabilities, thresholding at 0.5 for binary labels. Datasets are divided into training and test sets, with adjustable parameters—noise level, sample size, and shape type—varied to explore model behavior. This setup, consistent across experiments with minor tweaks (e.g., dimensionality), evaluates the model’s ability to detect signals under controlled conditions.

\begin{remark}[Parameter Tuning]
We tuned the cyclic-shift SKPD model’s hyperparameters to optimize classification performance on a $128 \times 128$ butterfly signal dataset, constructed as described above. Using the decomposition $\mathbf{C} = \sum_r \mathbf{A}{j,r} \otimes \mathbf{B}{j,r}$, we sequentially adjusted: (1) $\alpha \in [0, 1]$ in \eqref{eq:skpd-penalized-loss} (elastic net balance for $\mathbf{B}$), with $\lambda_a = \lambda_b = 0.1$ fixed, in steps of 0.1; (2) Using the best $\alpha = 0.2$ from the experiment, we then tuned $\lambda_a, \lambda_b \in \{0.001, 0.01, 0.1, 0.5, 1, 5, 10\}$, using grid search with the selected $\alpha$, evaluating performance via AUC and accuracy on a test set.
\end{remark}

\begin{figure}[ht]
    \centering
    \includegraphics[width=0.65\linewidth]{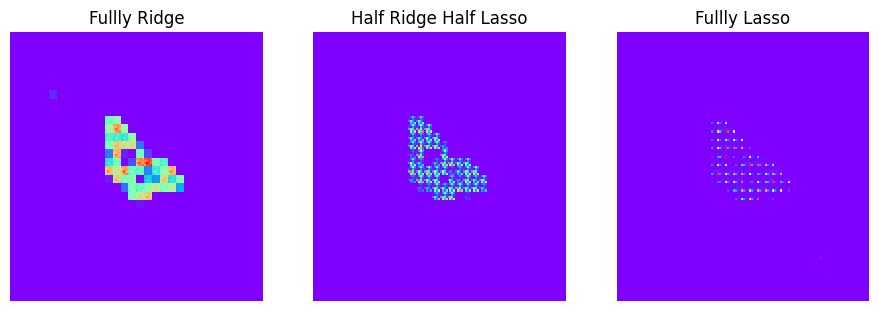}
    \caption{An illustration of estimated coefficients $\hat C \in \mathbb R^{128\times 128}$ from different regularization parameter on $\mathbf B_r$ matrix update.}
    \label{fig: diff reglarization on B}
\end{figure}

\begin{figure}[ht]
    \centering
    \includegraphics[width=0.65\linewidth]{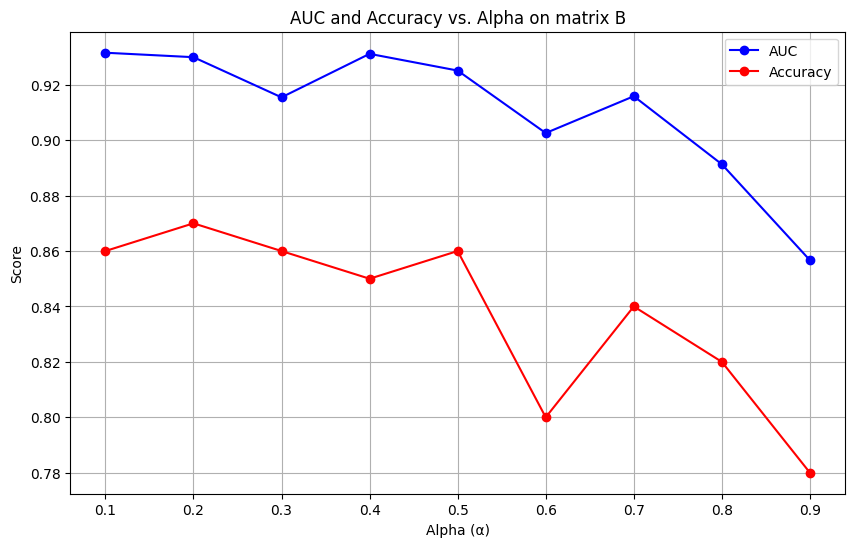}
    \caption{Model performance with varying $\alpha$ for $\mathbf{B}$'s update.}
    \label{fig: auc on diff alpha}
\end{figure}

The tuning results, illustrated in Figure~\ref{fig: auc on diff alpha}, show that $\mathbf{B}$, with its smaller dimensions ($4 \times 4 \times 1$, totaling 16 elements), benefits from a ridge-dominated regularization ($\alpha = 0.2$). This preserves the smooth, non-sparse local patterns critical for reconstructing the structured butterfly signal, as excessive sparsity (higher $\alpha$) disrupts these patterns, leading to a performance drop (AUC decreases from 0.9299 at $\alpha = 0.2$ to 0.8494 at $\alpha = 1.0$). Figure~\ref{fig: diff reglarization on B} further demonstrates this: The fully ridge ($\alpha = 0$) captures broader signal regions but over-smooths, the fully Lasso ($\alpha = 1$) enforces excessive sparsity and loses circular structures, while the elastic net ($\alpha = 0.5$) balances sparsity and smoothness, accurately preserving key features like rings and smaller circles.

\begin{figure}[ht]
    \centering
    \includegraphics[width=0.65\linewidth]{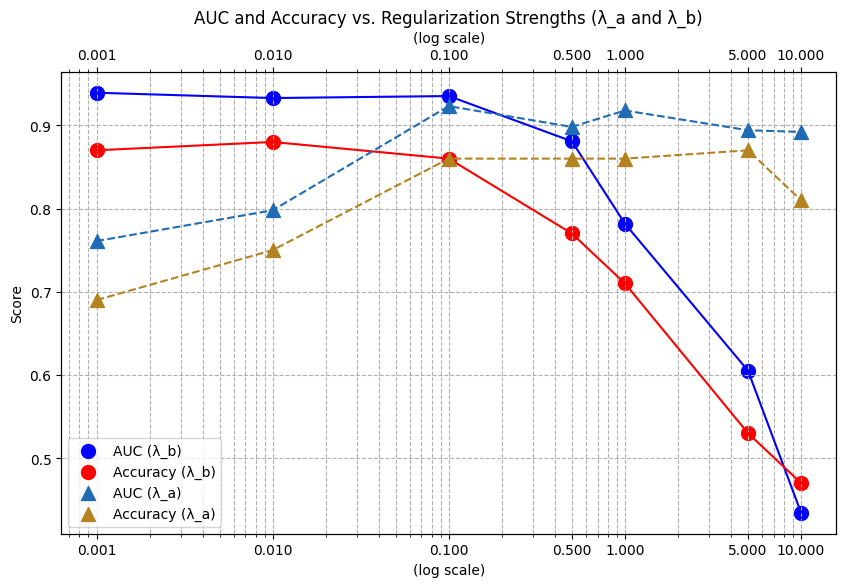}
    \caption{Model performance with different regularization strengths $\lambda_a$ and $\lambda_b$.}
    \label{fig: regularization strengths}
\end{figure}

The subsequent tuning of $\lambda_b$ and $\lambda_a$, shown in Figure~\ref{fig: regularization strengths}, reveals a pattern where the optimal $\lambda_b = 0.001$ (AUC 0.9392, accuracy 0.8700) is significantly smaller than the optimal $\lambda_a = 0.1$ (AUC 0.9231, accuracy 0.8600). This pattern arises due to the size disparity between the matrices: $\mathbf{A}$, with dimensions $32 \times 32 \times 1$ (1024 elements), has more degrees of freedom and a higher risk of overfitting, necessitating a larger $\lambda_a$ to enforce sparsity via L1 regularization and focus on relevant features. Conversely, $\mathbf{B}$’s smaller size means it has fewer parameters to overfit, so a smaller $\lambda_b$ suffices to prevent over-regularization while retaining its structural role. These optimal values may vary across datasets with different signal structures, noise levels, or dimensions, underscoring the need for dataset-specific tuning.

\begin{remark}[Patch size effect on $\mathbf A_r$]
We investigate how the patch size of $\mathbf{A}_r$ affects the SKPD model’s ability to reconstruct the spatial structure of a $128\times128$ butterfly signal, a critical factor for capturing MRI-like patterns. We tested $\mathbf{A}_r$ sizes of $64\times 64, 32\times 32, 16 \times 16, 8\times 8, 4\times 4$, and $2\times 2$, with corresponding $\mathbf{B}_r$ sizes of $2\times2$ to $64\times 64$, satisfying $p_1 \times d_1 = 128$, $p_2 \times d_2 = 128$. This partitions $\mathbf{C}$ into a grid of patches (e.g., size $4\times 4 \mathbf{A}_r$ yields 16 patches). Hyperparameters were set to $\alpha = 0.2$, with 5-fold cross-validation for evaluation.
\end{remark}

\begin{figure}[H]
    \centering
    \includegraphics[width=0.65\linewidth]{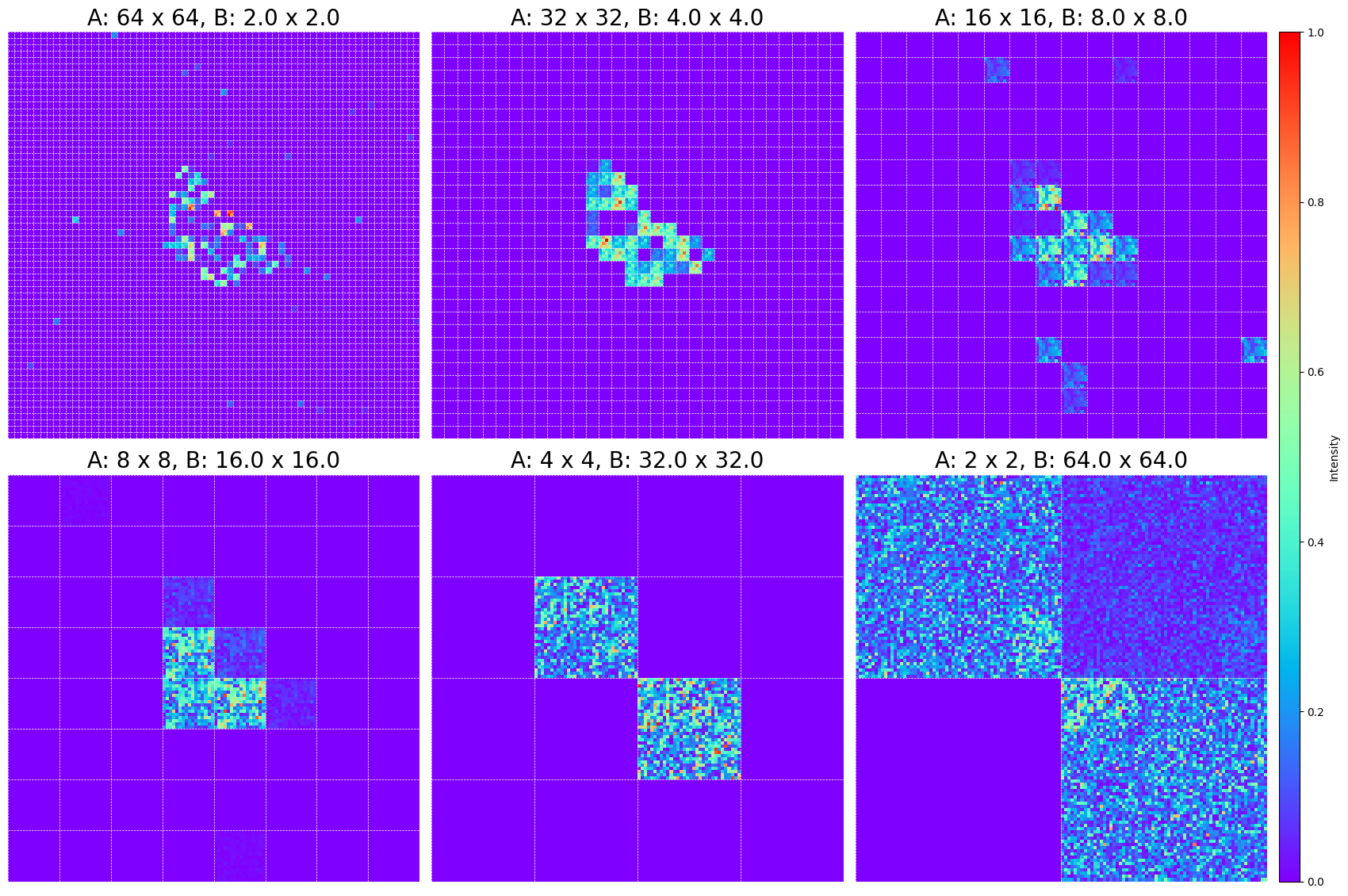}
    \caption{Estimated $\mathbf{C}$ matrices illustrating the impact of varying patch sizes of $\mathbf{A}_r$}
    \label{fig:grid_size_heatmaps}
\end{figure}

Table~\ref{tab:diff grid} shows that larger patch sizes lead to better model performance, with the $32 \times 32$ patch for $\mathbf{A}_r$ achieving the highest scores (mean AUC 0.9875 (SD: 0.0072), mean accuracy 0.9540 (SD: 0.0102)). In contrast, smaller patch sizes like $4 \times 4$ and $2 \times 2$ see significant drops in performance, with accuracies of 0.7580 (SD: 0.0194 and 0.6800 (SD: 0.0410), respectively. Figure~\ref{fig:grid_size_heatmaps} illustrates this trend: larger $\mathbf{A}_r$ patches (e.g., $32 \times 32$) effectively capture the butterfly’s global structure, while smaller sizes (e.g., $4 \times 4$, which splits $\mathbf{C}$ into 16 patches) tend to overfit to noise, fragmenting the underlying signal. This degradation occurs because large $\mathbf{A}_r$ patches model broader spatial dependencies, whereas smaller patches primarily respond to local noise. An optimal range of $32 \times 32$ to $64 \times 64$ balances global and local spatial features, providing practical guidance for patch size selection.

\begin{table}[ht]
    \normalsize                
\centering
    \begin{tabular}{lccc}
        \toprule
        Matrix Size of ($\mathbf{A}_r$) & Matrix Size of ($\mathbf{B}_r$)& Mean Accuracy (SD) & Mean AUC (SD) \\
        \midrule
        $64 \times 64$ & $2 \times 2$ & $0.8780 (0.0172)$ &$ 0.9502 (0.0086)$ \\
        $32 \times 32$ & $4 \times 4$ & $0.9540 (0.0102$) & $0.9875 (0.0072)$ \\
        $16 \times 16$ & $8 \times 8$ & $0.7820 (0.0697)$ & $0.8723 (0.0719)$ \\
        $8 \times 8$ &  $16 \times 16$& $0.8060 (0.0280)$ & $0.8932 (0.0320)$ \\
        $4 \times 4$ & $32 \times 32$ & $0.7580 (0.0194)$ & $0.8502 (0.0186)$ \\
        $2 \times 2$ & $64 \times 64$ & $0.6800 (0.0410)$ & $0.7495 (0.0252)$ \\
        \bottomrule
    \end{tabular}
    \caption{Performance metrics--mean and standard deviation (SD) of accuracy and AUC for different patch sizes of $\mathbf{A}_r$.}
    \label{tab:diff grid}
\end{table}

\subsubsection{Noise Robustness and Cyclic-Shift Impact}
MRI data often contains noise that can obscure disease signals, so we test the SKPD model’s robustness by varying noise levels and assess whether cyclic-shifting improves performance. Using the circles shape dataset (with butterfly results in figures), we compared two models---Cyclic-Shift SKPD and Non-Cyclic-Shift SKPD---across Gaussian noise levels $\sigma \in \{1.0, 5.0, 10.0, 15.0\}$, using 5-fold cross-validation. This range spans low to high noise, aiming to find the model’s breaking point.

\begin{figure}[ht]
    \centering
    \includegraphics[width=.8\linewidth]{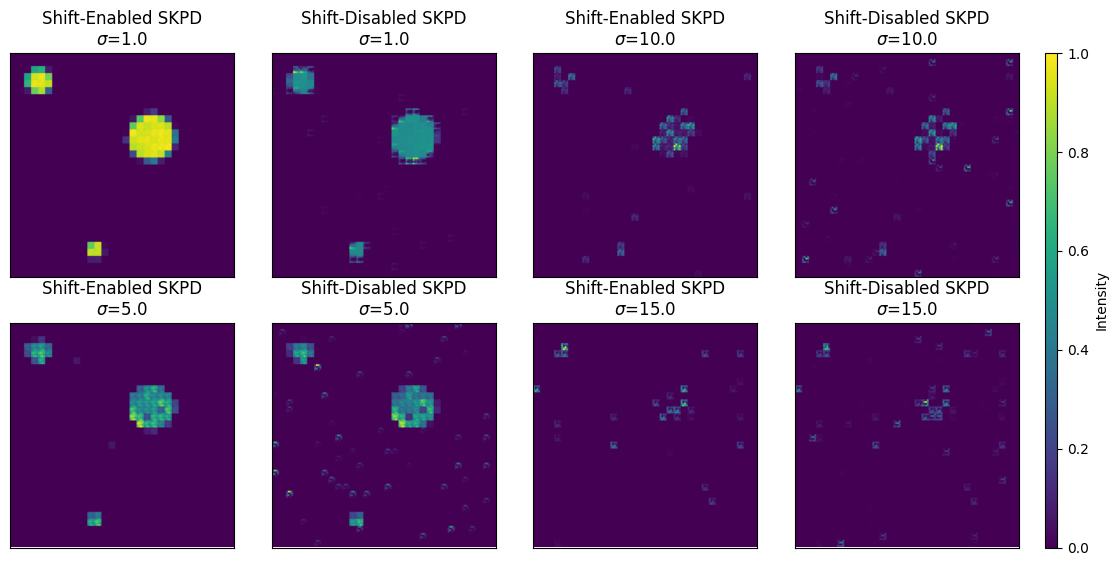}
    \caption{Estimated $\mathbf{C}$ matrices for the circles shape across varying noise levels ($\sigma$) and models.}
    \label{fig:noise_circles}
\end{figure}

\begin{figure}[ht]
    \centering
    \includegraphics[width=.8\linewidth]{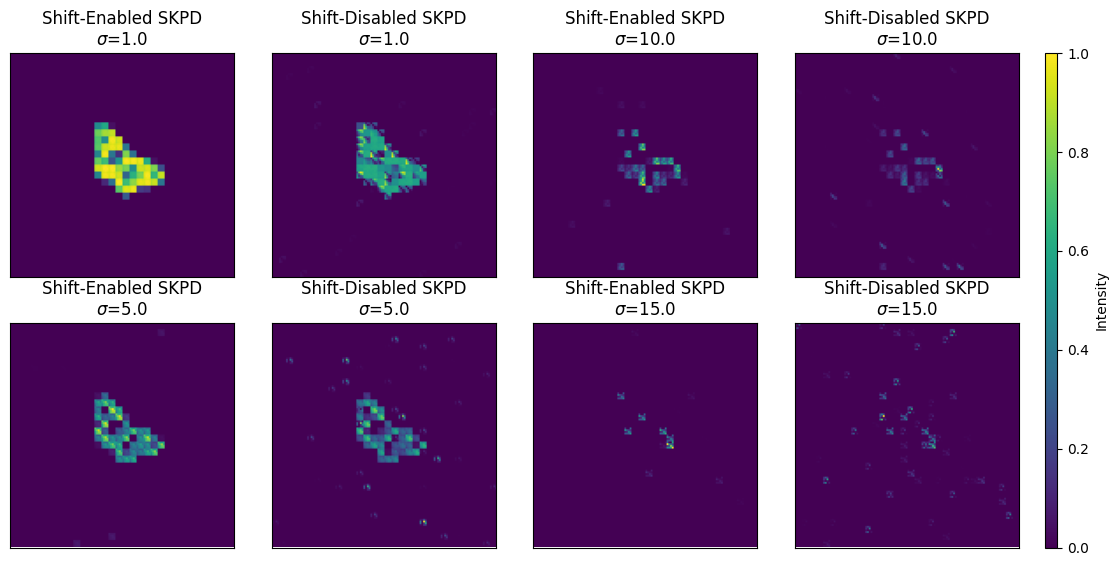}
    \caption{Estimated $\mathbf{C}$ matrices for the "butterfly" shape across varying noise levels ($\sigma$) and models.}
    \label{fig:noise_butterfly}
\end{figure}

Table~\ref{tab:noise_study} shows that the model maintains strong performance under low to moderate noise levels ($\sigma \leq 5.0$), with near-perfect mean AUC 0.9997 (SD: 0.0005) for Cyclic-Shift SKPD at $\sigma = 5.0$. At $\sigma = 10.0$, mean classification accuracy begins to drop to 0.8250 (SD: 0.0411), and mean AUC falls to 0.9138 (SD: 0.0298). By $\sigma = 15.0$, the model performance deteriorates substantially, with mean AUC reduced to 0.5861 (SD: 0.0890), indicating that noise has significantly disrupted the learned patterns. Figures~\ref{fig:noise_circles} and \ref{fig:noise_butterfly} confirm this visually: at $\sigma = 15.0$, distinct patterns such as circles are lost to noise artifacts. These results suggest a practical upper bound around $\sigma \approx 10.0$ for reliable classification using SKPD.

Cyclic-shifting provides modest but consistent gains, especially under higher noise. For instance, at $\sigma = 10.0$, Cyclic-Shift SKPD achieves a mean AUC of 0.9138 (SD: 0.0298) compared to 0.8646 (SD: 0.0303) for the Non-Cyclic-Shift SKPD. This indicates that shifting helps to realign spatial patterns degraded by noise. These findings underscore the robustness of the SKPD framework in handling realistic noise levels common in MRI data, while also demonstrating that shift-invariance adds resilience under challenging conditions.

\begin{table}[ht]
    \normalsize                
\centering
    \caption{Performance metrics for the circles shape across different noise levels ($\sigma$) and models.}
    \label{tab:noise_study}
    \begin{tabular}{|c|c|c|c|}
    \hline
    $\sigma$ & Model & Mean Accuracy (SD) & Mean AUC (SD) \\
    \hline
    1.0 & Cyclic-Shift SKPD & $1.0000 (0.0000)$ & $1.0000 (0.0000)$ \\
    1.0 & Non-Cyclic-Shift SKPD & $1.0000 (0.0000)$ & $1.0000 (0.0000)$ \\
    5.0 & Cyclic-Shift SKPD & $0.9925 (0.0100)$ & $0.9997 (0.0005)$ \\
    5.0 & Non-Cyclic-Shift SKPD & $0.9875 (0.0112)$ & $0.9987 (0.0020)$ \\
    10.0 & Cyclic-Shift SKPD & $0.8250 (0.0411)$ & $0.9138 (0.0298)$ \\
    10.0 & Non-Cyclic-Shift SKPD & $0.7950 (0.0400)$ & $0.8646 (0.0303)$ \\
    15.0 & Cyclic-Shift SKPD & $0.5850 (0.0567)$ & $0.5861 (0.0890)$ \\
    15.0 & Non-Cyclic-Shift SKPD & $0.5425 (0.0645)$ & $0.5675 (0.0552)$ \\
    \hline
    \end{tabular}
\end{table}

\begin{table}[ht]
    \normalsize                
\centering
    \caption{Performance metrics for the butterfly shape across different noise levels ($\sigma$) and models.}
    \label{tab:noise_study_butterfly}
    \begin{tabular}{|c|c|c|c|}
    \hline
    $\sigma$ & Model & Mean Accuracy (SD) & Mean AUC (SD) \\
    \hline
    1.0 & Cyclic-Shift SKPD & $1.0000 (0.0000)$ & $1.0000 (0.0000)$ \\
    1.0 & Non-Cyclic-Shift SKPD & $1.0000 (0.0000)$ & $1.0000 (0.0000)$ \\
    5.0 & Cyclic-Shift SKPD & $0.9900 (0.0050)$ & $0.9994 (0.0008)$ \\
    5.0 & Non-Cyclic-Shift SKPD & $0.9775 (0.0184)$ & $0.9981 (0.0014)$ \\
    10.0 & Cyclic-Shift SKPD & $0.7800 (0.0423)$ & $0.8782 (0.0392)$ \\
    10.0 & Non-Cyclic-Shift SKPD & $0.7275 (0.0267)$ & $0.8074 (0.0270)$ \\
    15.0 & Cyclic-Shift SKPD & $0.5825 (0.0650)$ & $0.6204 (0.0896)$ \\
    15.0 & Non-Cyclic-Shift SKPD & $0.5625 (0.0403)$ & $0.5892 (0.0304)$ \\
    \hline
    \end{tabular}
\end{table}

\subsection{Tensor-Image Simulation}
In the 3D tensor-image simulation experiment, a simulated dataset was generated to distinguish between two classes using noisy tensor structures. The goal was to construct tensor data with distinct spatial patterns while introducing noise to simulate real-world MRI-like variability. The upper two plots in the provided image illustrate the simulated data for each class, while the lower plot presents the Kronecker product reconstruction, highlighting the model's ability to capture key differences between the two classes. The lower part of the spherical structures represents the primary difference between the two simulated classes.

The dataset consists of 1,000 samples, with 500 samples per class. Each sample was generated by adding Gaussian noise with a standard deviation of 1 to a structured base tensor specific to each class. Class 1 samples were based on one base tensor, while Class 0 samples used a slightly modified version with only one ball, ensuring class-distinct patterns.

The high classification accuracy of 95\% and AUC of 0.9713 are due to the clear structural differences between the two classes, despite the added noise. The distinguishing features embedded in the base tensors allow the model to learn separable representations. Furthermore, the Kronecker product reconstruction effectively captures these differences, particularly in the lower region of the spherical structures, enabling the classifier to leverage these key characteristics.

Since the base tensors are designed to have inherent class-specific structures, and the noise level (standard deviation of 1) is relatively low, the model can consistently recognize patterns and make perfect predictions.
\begin{figure}[H]
    \centering
    \includegraphics[width=1\linewidth]{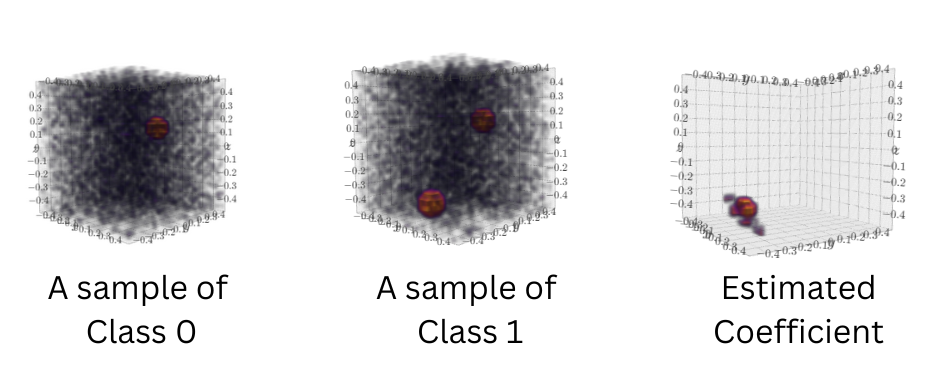}
    \caption{To demostrate how different class in the simulation 3D tensor dataset looks like, the $\sigma$ value showing here is lower for better visualization explain. The estimated coeffecient tensor perfectly capture the lower ball is the difference between two classes and capture most of the shape of the ball.}
\end{figure}

\section{Applications to OASIS-1 and ADNI-1 MRI data}
\label{sec:mri_application}

We assess the cyclic–shift SKPD classifier on two public neuro-imaging
cohorts that differ in sample size, acquisition protocol and
pre-processing pipeline.

\textbf{OASIS-1.}  
The Open Access Series of Imaging Studies provides three to four
T1-weighted scans for each of 416 participants aged 18–96.  
One hundred participants aged over 60 have a Clinical Dementia Rating
(CDR) above~0 and are labelled as probable Alzheimer’s disease (AD)
cases.  Volumes of size $176 \times 208 \times 176$ are
shape- and intensity-normalised to $64^{3}$ voxels.  Covariates
include age, sex, total intracranial volume, normalised whole-brain
volume and an atlas-scaling factor; continuous variables are
standardised.

\textbf{ADNI-1.}  
From the ADNI archive we use 246 month-12 scans acquired on Siemens
1.5-T systems.  After identical normalisation each volume is
$64^{3}$.  The cohort comprises 77 cognitively normal (CN), 125
mild-cognitive-impairment (MCI) and 44 AD subjects; for consistency
with OASIS-1 we dichotomise the diagnosis into AD
(CDR\,$>\!0$) versus non-AD.  Metadata contain age (56–91), sex and
visit information.

\subsubsection*{Two-stage modelling pipeline}

To exploit the spatial interpretability of the 3-D coefficient tensor
we adopt a two-stage procedure.  Stage 1 fits the cyclic-shift SKPD
model to the full $64^{3}$ tensors together with subject-level
covariates,
\[
\Pr\{Y_i=1\mid\mathbf{X}_i,\mathbf{z}_i\}
  =\mathrm{logit}^{-1}\!\bigl\langle\widehat{\mathbf{C}},
          \mathbf{X}_i\bigr\rangle
     +\mathbf{z}_i^{\!\top}\widehat{\boldsymbol\gamma},
\]
yielding a structured coefficient array
$\widehat{\mathbf{C}}$ (Figure~\ref{fig:oasis_3d}).  
For every anatomical plane we sum
$|\widehat{\mathbf{B}}|$ across slices and retain the slice with
largest total magnitude.

Stage 2 constructs three data sets—axial, coronal and sagittal—based
on these most informative slices and fits a separate 2-D
cyclic-shift SKPD model to each.  This view-specific training
enhances discrimination while allowing angle-by-angle interpretation.
Unlike conventional pipelines that use fixed median slices or expert
selection, the procedure is entirely data-driven;  
Algorithm~\ref{alg:mri_pipeline} summarises the steps.

\subsubsection*{Two-stage modeling pipeline}

\begin{figure}[ht]
    \centering
    \includegraphics[width=0.85\linewidth]{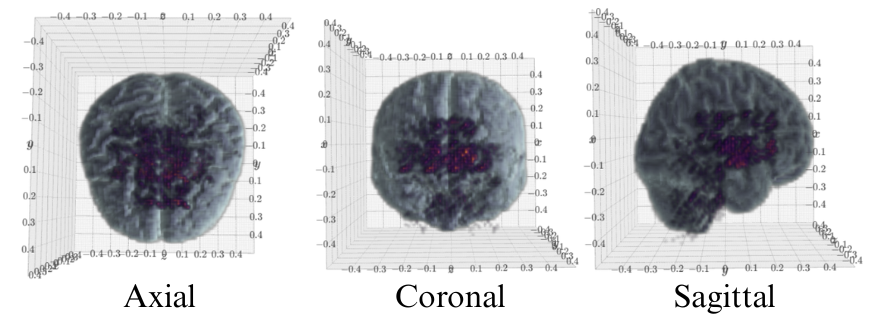}
    \caption{Different angles of an MRI with the estimated coefficient pointing out the important signal difference between AD patients and non AD patients on the OASIS-1 Dataset}
    \label{fig:oasis_3d}
\end{figure}

Let $\mathcal{X}_i \in \mathbb{R}^{64 \times 64 \times 64}$ denote the MRI tensor for subject $i$, and let $\mathbf{z}_i \in \mathbb{R}^5$ represent the vector of standardized demographics. In Stage 1, we fit the full 3D cyclic-shift SKPD logistic model:
\[
\Pr(Y_i = 1 \mid \mathbf{X}_i, \mathbf{z}_i) = \mathrm{logit}^{-1}\left\{\langle \mathbf{C}, \mathbf{X}_i \rangle + \mathbf{z}_i^{\top} \boldsymbol{\gamma} \right\},
\]
producing a coefficient tensor $\widehat{\mathbf{C}}$ whose voxel-wise magnitudes highlight discriminative regions (Figure~\ref{fig:oasis_3d}). For each anatomical axis (axial, coronal, sagittal), we sum $|\widehat{\mathbf{C}}|$ across slices and retain the one with maximal total intensity.

Stage 2 trains three independent 2D cyclic-shift SKPD models—one per selected slice—enabling angle-specific interpretations without re-running the full 3D optimization. The complete procedure is outlined in Algorithm~\ref{alg:mri_pipeline}; the algorithmic format is used solely for clarity and does not imply a different estimation strategy.

\begin{algorithm}[H]
\caption{Two-stage data-driven slice selection and multi-view 2D modeling}
\label{alg:mri_pipeline}
\begin{algorithmic}[1]
\State \textbf{Stage 1.} Fit the 3D cyclic-shift SKPD model to $(\mathbf{X}_i, \mathbf{z}_i, Y_i)$ and save $\widehat{\mathbf{C}}$.
\For{each anatomical plane $p \in \{\text{axial}, \text{coronal}, \text{sagittal}\}$}
      \State Compute $s_{p,k} = \sum_{(j,\ell)} \bigl| \widehat{\mathbf{C}}_{p,k}(j,\ell) \bigr|$ for every slice $k$.
      \State Select $k^\star_p = \arg\max_k s_{p,k}$ and extract the corresponding image slice from every subject.
      \State Fit a 2D cyclic-shift SKPD model to the selected slice and covariates $\mathbf{z}_i$.
\EndFor
\State \textbf{Return} three fitted 2D models and their coefficient maps.
\end{algorithmic}
\end{algorithm}

\subsubsection*{Predictive performance and interpretation}

\begin{figure}[ht]
    \centering
    \includegraphics[width=0.85\linewidth]{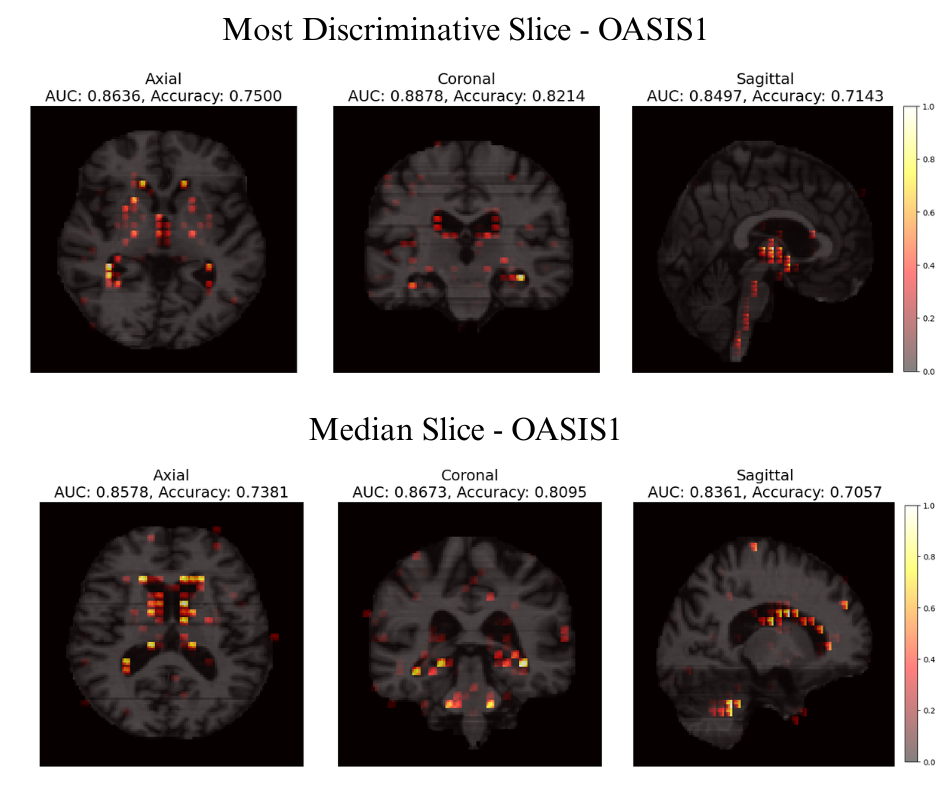}
    \caption{Overlay of MRI slices of the most discriminative slices(Axial: 74/176, Coronal: 98/208, Sagittal: 85/176) and median slices with estimate coefficients for Axial, Coronal, and Sagittal views on the OASIS-1 Dataset}
    \label{fig:mri_2d_best}
\end{figure}

Cross-validated results for OASIS-1 yield a mean AUC of 0.8784 (SD: 0.0253) and mean accuracy of 0.8072 (SD: 0.0300), surpassing the standard (non-shifted) SKPD baseline (AUC 0.8658, SD: 0.0217; accuracy 0.7842, SD: 0.0224). Visualizations of brain slices from an AD patient, overlaid with the estimated 2D model coefficients, reveal that the selected most discriminative slices (Axial: 74/176, Coronal: 98/208, Sagittal: 85/176) concentrate attention on central brain regions—most notably the hippocampus and medial temporal lobe (Figure~\ref{fig:mri_2d_best}). In contrast, median slices often highlight peripheral cortical areas, which may reflect brain shrinkage associated with normal aging rather than AD-specific pathology, potentially diluting the specificity of the signal. By emphasizing central regions more strongly associated with AD pathology, the data-driven slice selection enhances both predictive accuracy and biological plausibility compared to heuristics like median slicing.

\begin{figure}[ht]
    \centering
    \includegraphics[width=0.85\linewidth]{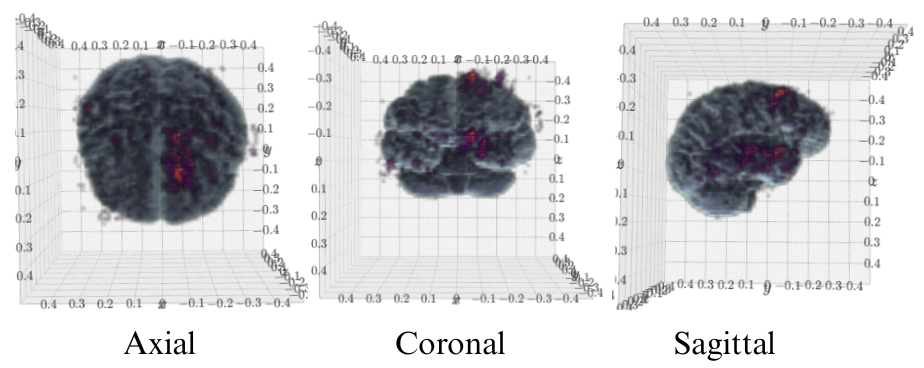}
    \caption{Different angles of an MRI with the estimated coefficient pointing out the important signal difference between AD patients and non AD patients on the ADNI1 Dataset}
    \label{fig:mri_adni_3d}
\end{figure}

\begin{figure}[ht]
    \centering
    \includegraphics[width=0.85\linewidth]{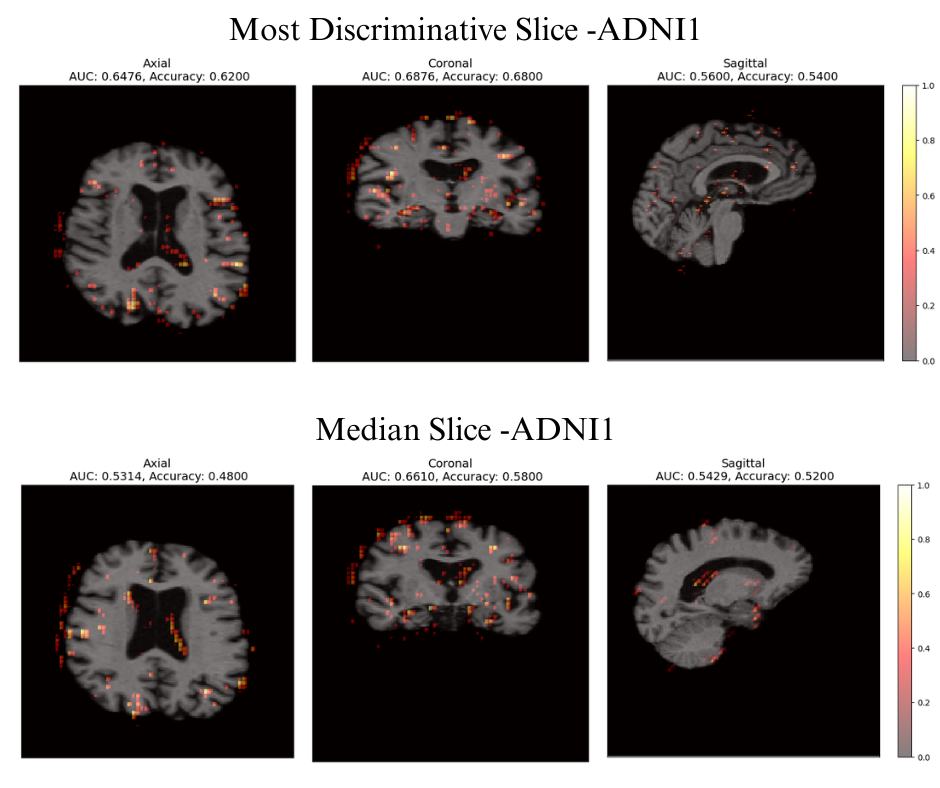}
    \caption{Overlay of MRI slices of the most discriminative slices(Axial: 65/192, Coronal: 100/192, Sagittal: 83/180)/median slices with estimate coefficients for Axial, Coronal, and Sagittal views on the ADNI1 Dataset}
    \label{fig:mri_adni_2d}
\end{figure}

ADNI-1 poses a harder problem because non-brain tissue cannot be fully removed by the standard pipeline. Nevertheless, the cyclic-shift SKPD attains a mean AUC of 0.6033 (SD: 0.0591) and accuracy of 0.5792 (SD: 0.0496), on par with recent deep-learning alternatives. Crucially, its coefficient maps still concentrate on the hippocampal formation (Figures~\ref{fig:mri_adni_3d}–\ref{fig:mri_adni_2d}), indicating that the model remains interpretable even when overall classification is challenging.

\subsubsection*{Comparison with existing SKPD imaging analyses}

\citep{wu2023_sparse} applied SKPD to the 3D MRI images with continuous responses and located sparse (one or two) signal-bearing blocks.  By contrast, the present logistic-tensor extension detects sparse voxel-level signals directly—without post-hoc thresholding—and operates successfully on binary disease status.  The marked concentration of coefficients within recognized AD-related structures therefore illustrates the gain in anatomical resolution afforded by the spatial–shift mechanism.

Overall, the real-data experiments corroborate the simulation evidence that cyclic-shifting stabilizes inference and that data-driven slice selection delivers clinically meaningful localization without manual input.

The 3D cyclic-shifted SKPD logistic models achieving the cross validated mean AUC of 0.8776 (SD: 0.0253) and mean accuracy 0.8069 (SD: 0.0299). Which is better than the standard SKPD model with a mean AUC 0.8649 (SD:0.0208) and mean accuracy 0.7836 (SD: 0.0221). Visualizations of brain slices from an AD patient, overlaid with the estimated model coefficients, reveal that the selected most discriminative slices concentrate attention on central brain regions—most notably the hippocampus and medial temporal lobe. In contrast, median slices often highlight edge areas, which may reflect brain shrinkage; while such peripheral atrophy can occur in AD, it is also commonly associated with normal aging, potentially diluting the specificity of the signal.

By emphasizing central regions more strongly associated with AD pathology, the best-slice selection aligns closely with established clinical understanding. This demonstrates that our data-driven approach not only improves model performance, but also enhances interpretability and clinical relevance compared to heuristics like median slicing.

Though the model performance on the ADNI1 dataset only achieves a cross-validated mean AUC 0.6027(SD: 0.0586) and mean accuracy 0.5793(SD: 0.0485) due to the challenge of MRI image process to get rid of the non-brain tissue. The visualization of the estimated coefficient still highlights the central brain regions, most notably the hippocampus and medial temporal lobe.

\section{Discussion and Conclusion}
\label{sec:discussion}

This paper introduces a cyclic-shift sparse Kronecker product decomposition (SKPD) classifier for high-dimensional tensor covariates. By integrating a low-rank  factorization with a computationally efficient cyclic-shift augmentation, the method achieves three key objectives: interpretability via explicit voxel-level coefficient maps, robustness to local mis-registration of anatomical structures, and scalability to full-resolution three-dimensional images.

Simulation studies show that cyclic-shifting improves signal recovery in moderate- to high-noise settings. They also provide a practical guideline for selecting the spatial factor $\mathbf{A}$ and signal factor $\mathbf{B}$: the dimension of $\mathbf{A}$ should be significantly larger than that of $\mathbf{B}$ but below the full image resolution, balancing fidelity to global spatial structure with overfitting control. These findings align with the non-asymptotic error bounds in Section~\ref{sec:theory}.
Analysis of real MRI data in Section~\ref{sec: Simulation experiments} demonstrates that the classifier accurately identifies clinically relevant brain regions, notably the hippocampus and medial temporal lobe, while delivering strong predictive performance on the OASIS-1 and ADNI-1 cohorts.

Several directions for future research are clear. First, adaptive dimension selection for the decomposition could enhance portability across imaging protocols and resolutions. Second, robust noise regularization, potentially via hybrid deep learning and statistical denoising, may improve performance in highly noisy acquisitions. Third, extending the framework to multi-class clinical phenotypes would support applications to progressive stages of cognitive decline. Finally, sharper theoretical results, such as convergence rates under correlated tensor designs, could further elucidate the regularization effects of the cyclic-shift operator.

In conclusion, the cyclic-shift SKPD tensor-logistic model offers an interpretable, theoretically grounded, and computationally efficient approach to tensor-based classification. Its strong performance on synthetic and real MRI data, combined with clear opportunities for methodological advances, establishes a solid foundation for future statistical and clinical innovations.

\section*{Acknowledgments}
We gratefully acknowledge the support of the National Science Foundation through grants (Hsin-Hsiung Huang: DMS-1924792, DMS-2318925 and Teng Zhang: CNS-1818500).  
The authors are grateful to the Editor, Associate Editor, and the anonymous referees for their insightful comments, which have substantially improved the manuscript.

\section*{Data availability}
The MRI datasets analyzed here are distributed by the Open Access Series of Imaging Studies (OASIS) and the Alzheimer’s Disease Neuroimaging Initiative (ADNI).  
Access requires registration and acceptance of each repository’s data-use agreement.

\noindent
\textit{OASIS-1:} \url{https://sites.wustl.edu/oasisbrains/} \\
\textit{ADNI-1:} \url{https://adni.loni.usc.edu/}

\noindent
A Python package that implements the cyclic–shift SKPD models used in this study is openly available at  
\url{https://github.com/JackieChenYH/CycVoxSKPD}.

\bibliography{csskpd}

\end{document}